# Transition from semiconducting to metallic-like conducting and weak antilocalization effect in single crystals of LuPtSb

Zhipeng Hou[1,2], Yue Wang[1], Guizhou Xu[1], Xiaoming Zhang[1], Enke Liu[1], Wenquan Wang[2], Zhongyuan Liu[3], Xuekui Xi[1], Wenhong Wang[1,a)], Guangheng Wu[1]

[1]State Key Laboratory for Magnetism, Beijing National Laboratory for Condensed Matter Physics, Institute of Physics, Chinese Academy of Sciences, Beijing 100190, China,
[2]College of Physics, Jilin University, Changchun 130023, China,
[3]State Key Laboratory of Metastable Material Sciences and Technology, Yanshan University, Qinhuangdao 066004, China,

## Abstract

High quality half-Heusler single crystals of LuPtSb have been synthesized by a Pb flux method. The temperature dependent resistivity and Hall effects indicate that the LuPtSb crystal is a $p$-type gapless semiconductor showing a transition from semiconducting to metallic conducting at 150 K. Moreover, a weakly temperature-dependent positive magnetoresistance (MR) as large as 109 % and high carrier mobility up to 2950 cm$^2$/Vs are experimentally observed at temperatures below 150 K. The low-field MR data shows evidence for weak antilocalization (WAL) effect at temperatures even up to 150 K. Analysis of the temperature and angle dependent magnetoconductance manifests that the WAL effect originates from the bulk contribution owing to the strong spin-orbital coupling.

Correspondence and requests for materials should be addressed to W.H.W (email: wenhong.wang@iphy.ac.cn).

The heavy ternary half-Heusler compounds with 1:1:1 composition and crystallize with the MgAgAs-type structure (space group F-43m), have recently drawn much attention due to their rich physics properties, such as topological insulators,[1-4] superconductivity,[5-8] liner magnetoresistance (MR),[9] ultrahigh electron mobility,[10] and heavy fermion behavior.[11] These multifunctional characteristics of half-Heusler compounds provide a very promising platform for both fundamental research and potential applications. Among the most studied half-Heusler compounds, LuPtSb is of interest as it features a weak inverted band structure and a small lattice distortion may easily open up the bulk band gap at Fermi level.[1, 12] Shekhar *et al* have investigated transport properties of LuPtSb and found a non-saturating linear MR behavior below 20 K, which is of great significance to both the quantum magnetotransport properties and magnetic sensor applications as well.[9] More recently, Sahil *et al* have successfully grown LuPtSb single crystal thin films on relaxed lattice-matched $Al_{0.1}In_{0.9}Sb$ buffer layers on GaAs substrates by molecular beam epitaxy (MBE) method.[13] Both scanning tunneling spectroscopy (STS) and photoemission measurements studies suggest that the position of the Fermi level locates in the valence band, confirming the results of recent band structure calculation. In the current work, we fabricated the high-quality bulk single crystals of LuPtSb, and reported the results of structural, magnetic and electrical transport properties. Transport investigations have revealed LuPtSb is a gapless semiconductor showing a crossover from semiconducting to metallic-like conducting at 150 K, a positive MR as large as 109 % and high carrier mobility up to 2950 $cm^2$/Vs at 2 K. More intriguingly, a clear weak antilocalization (WAL) effect is experimentally observed at temperatures even up to 150 K.

High-quality single crystals of LuPtSb were synthesized by a Pb flux method as described in our previous reports.[5, 14] The crystal structure was checked by powder X-ray diffraction (XRD) measurement which was carried out on crushed single crystals using a Rigaku X-ray diffractometer with Cu-*K*α radiation. The single-crystal orientation was checked by a standard Laue diffraction technique. The transport measurements were performed on a Quantum Design physical properties measurement system (PPMS) in the temperature range from 300 to 2 K. The longitudinal MR was measured with the magnetic field perpendicular to the electrical current direction. The Hall effect was measured by rotating the crystal by 180 ° in a magnetic field of 5 T. For both the MR and Hall effect measurements, the current and magnetic field correspond to the in-plane and out-of-plane directions of the (111) plane of the sample.

Figure 1 (a) shows a typical photographic image of single crystal of LuPtSb with a dimension of approximately $1.2 \times 1.2 \times 0.6$ mm$^3$. It can be seen that the crystal exhibits mirror-like surfaces and is robust in the air. To examine the quality of the single crystals, a standard X-ray Laue back-reflections technique was employed across the hexagon surface facing us. As shown in Figure 1 (b), the Laue image shows clear six-fold symmetry, strongly suggesting that the sample was a high-quality single crystal with the well-developed (111) facet. To further determine the crystal structure, the powder XRD measurement was carried out and the corresponding pattern is presented in Figure 1 (c). It is noticeable that all the diffraction peaks can be well indexed to the MgAgAs-type crystal structure (space group *F-43m*), with a refined lattice parameter $a = 6.4677$ Å, which is in reasonable agreement with that of polycrystalline sample reported by Shekhar *et al*.[9]

Figure 2 (a) shows the temperature dependence of longitudinal resistivity $\rho_{xx}$ under a zero magnetic field. At 300 K, the single crystal sample exhibits a small resistivity $\rho_{xx} = 2.2$ μΩ m, which is nearly one order of magnitude lower than that of the polycrystalline one,[9] indicating the high-quality. Over the whole temperature range, the resistivity curve displays two different regimes. With decreasing the temperature from 300K, $\rho_{xx}$ firstly increases and reaches a maximum value of 4.9 μΩ m at 150 K, which is reminiscent of the semiconducting-like behavior ($d\rho/dT<0$). Below 150 K, it decreases in a metallic-like behavior ($d\rho/dT>0$). We should point out that, the semiconducting-to-metallic crossover is not extraordinary and commonly seen in the semimetal or gapless semiconductors that show small activation energies.[10, 15, 16] Moreover, the resistivity below 150 K is similar to the case in doped semiconductors, where some donor or acceptor levels are present due to the existence of point defects and/or little variation in the stoichiometry.[17]

To obtain the further insight of the transport properties, the Hall effect was measured. The inset of Figure 2 (b) presents the magnetic field dependence of the Hall resistivity $\rho_{xy}$ at various temperatures under a magnetic field up to 5 T. A liner positive Hall resistivity is observed over the whole temperature range, reflecting that only one type of hole charge carrier dominates the transport properties. The result coincides with the Hall measurements on the polycrystalline sample and also thin film.[9, 13] The estimated the carrier concentration *n* and Hall mobility $\mu_H$ and their temperature dependency are presented in Figure 2 (b) and (c), respectively. As shown in

Figure 2 (b), the carrier concentration ranges from $5.5 \times 10^{19}$ cm$^{-3}$ to $5.0 \times 10^{18}$ cm$^{-3}$ between 300 K and 2 K, which indicates the bulk single crystal of LuPtSb belongs to the low-charge-carrier semimetal or gapless semiconductor. By scanning the shape of the curve, the carrier concentration also displays a semiconductor-like (positive carrier concentration temperature dependence) and metallic-like (mostly carrier concentration temperature independence) behavior at the temperature ranges 300 K – 150 K and 150 K – 2 K, respectively, which agrees with the resistivity temperature dependency. On the other hand, we find that the Hall mobility is determined to be 525 cm$^2$/Vs at 300 K. With the decrease of temperature, it increases drastically to a high value of 2950 cm$^2$/Vs at 2 K, which is about 4 times larger than that of polycrystalline of LuPtSb (760 cm$^2$/Vs at 2 K) and comparable to that of polycrystalline of YPtSb (2500 cm$^2$/Vs at 2 K) and LuPdSb (1870 cm$^2$/Vs at 2 K). [9, 11] Usually, the temperature dependence of mobility is mainly affected by two types of scattering mechanisms: the ionized impurity scattering results in a positive temperature dependence of mobility, whereas the lattice vibration scattering leads to a negative result. In the case of the bulk single crystal of LuPtSb, the mobility firstly increases with a $T^{-2.5}$ behavior up to 150 K, and then follows a negligible temperature variation of $T^{-0.05}$ with further decreasing the temperature. The negative power law over the whole temperature suggests that the lattice vibration scattering dominates the temperature-dependent mobility.

Figure 3 (a) shows the magnetic field dependence of MR at various temperatures range from 300 K to 2 K with the magnetic field perpendicular to the current and also the sample (111) plane. To eliminate the effect from Hall resistance, the measurements were carried out from 8 T to -8 T and corresponding calculations were employed to obtain intrinsic MR. The MR value is defined as MR = [$\rho(H)$ - $\rho(0)$] / $\rho(0)$] × 100 %, where $\rho(H)$ and $\rho(0)$ are the resistivity at magnetic field $H$ and zero, respectively. At 300 K, a large positive MR of 57 % was obtained at 8 T. With the decrease of temperature, MR increases correspondingly and reaches a peak value of 109 % at 150 K. At temperatures below 150 K, MR is weakly dependent on the temperature and a value as large as 103% was achieved at 2 K and 8 T. In the case of polycrystalline LuPtSb, however, a quite small value of 2 % - 4 % was found below 20 K,[10] indicating that MR is drastically enhanced in the single crystal form. Moreover, it is found that MR exhibits a steep increase in the low magnetic field region and nearly reaches to saturation at 2 T (the values of MR$_{2T}$/ MR$_{8T}$ at 2 K, 50 K, 150 K are 0.86, 0.84, 0.83, respectively) below 150 K. The large and highly sensitive low-field

MR suggests that the half-Hesuler LuPtSb could potentially be for highly sensitive magnetic sensors.

The observed pronounced MR cusp at low fields can be associated with the weak antilocalization (WAL) effect, which is usually found in three-dimensional (3D) systems with strong spin-orbital coupling or two-dimensional (2D) surface states of topological insulators.[5,18-20] In order to investigate the origin of WAL in single crystal of LuPtSb, the Hikami-Larkin-Nagaoka (HLN) formula,[21]

$$\Delta G (H) = \frac{\alpha e^2}{2\pi^2 \hbar}[\psi(\frac{1}{2} + \frac{\hbar}{4eHL_\varphi}) - \ln(\frac{\hbar}{4eHL_\psi})]$$

, where $\psi$ is the digamma function, $L_\varphi$ is the phase coherence length and the prefactor $\alpha$ is experimentally found to be equal to -0.5 for each conductive channel in a traditional 2D electron system, is used to fit the magnetoconductance $\Delta G = G$ (H) - $G$ (0). Figure 3 (b) displays the temperature dependence of $\Delta G$ under a low magnetic field range from 0.2 T to -0.2 T between 2 K and 50 K. It is clearly evident that all the curves can be fitted well with the HLN model at the magnetic field range. The fitted values of $L_\varphi$ at different temperatures are presented in Figure 3 (c) and the inset shows the temperature dependence of $\alpha$. We observed that the fitting of $L_\varphi$ shows a $T^{-3.3}$ behavior and a $T^{0.1}$ behavior in the temperature regime 50 K-25 K and 25 K-2 K, respectively, which is similar to that of the single crystalline nanoflake devices of $Bi_{1.5}Sb_{0.5}Te_{1.8}Se_{1.2}$.[18] As shown in the inset of Fig. 3(c), the values of coefficient $\alpha$ are in the order of $10^5$, which are much larger than the theoretical 2D ones. Similar as the previous reported LuPdBi and SnTe,[5, 20] the large values of $\alpha$ suggest that the observed WAL behavior mainly originates from the 3D bulk contribution.

To further confirm the 3D-domainted WAL in the single crystal of LuPtSb, we have carried out the angle-dependent $\Delta G$ measurements at 2K under a maximum magnetic field up to 0.5 T. In Figure 4, we show the dependence of $\Delta G$ on the perpendicular component of the magnetic field $H$ $\sin\theta$, and $\theta = 90°$ means the magnetic field perpendicular to the magnetic field (see the inset). If WAL effect originates from the 2D surface states of topological insulators solely, the $\Delta G$ vs $H \sin\theta$ curves should overlap onto one curve. As shown in Figure 3 (b), however, the four curves deviate from each other in particular at high fields, confirming that the WAL effect in the bulk single crystal LuPtSb mainly results from the contribution of the 3D bulk spin-orbit coupling.

In conclusion, we have successfully synthesized the high-quality single crystal of LuPtSb by a Pb flux method. Experimental investigations of transport properties have demonstrated it a gapless semiconductor or semimetal with large positive MR (57 % - 109 %) and high Hall mobility (525 - 2950 cm$^2$/Vs) between 300 and 2 K. Compared with the polycrystalline crystal and thin film of LuPtSb, both the values of MR and mobility are enhanced significantly. Moreover, we have observed a clear WAL effect from the low magnetic field dependence of MR below 150 K. The angle dependence of magnetoconductivity and large values of $\alpha$ suggest that the WAL effect in bulk single crystal of LuPtSb does not originate from the possible 2D surface states but the contribution of strong 3D bulk spin-orbital coupling. In this regard, further experimental work will be necessary to detect the possible surface states in LuPtSb by doping or applying pressure to open the "closed" bulk band gap.

**Acknowledge**

This work is supported by the National Basic Research Program of China (973 Program 2012CB619405), National Natural Science Foundation of China (Grant Nos. 51171207 and 11474343), and Graduate Innovation Fund of Jilin University (No. 2014006).

**Figure captions:**

FIG. 1 (a) The photographic image of a typical single crystal of LuPtSb placed on a millimeter grid. (b) Laue diffraction pattern of single-crystal LuPtBi generated with the beam axis coincident with the [111] zone axis. (c) Observed (green circles) powder XRD patterns of crushed single crystals at room temperature and structural refinement results (red line). The differences between observed and calculated profiles are presented by the blue trace. The black segments show the expected diffraction peaks. The inset shows a structure view of conventional LuPtSb unit cell which has 4 number of formulary units.

FIG. 2 (a) Temperature dependence of longitudinal resistivity $\rho_{xx}$ under a zero magnetic field. (b) The variation of carrier concentration with temperature deduced from the Hall effect. The inset shows the liner behavior of Hall resistivity $\rho_{xy}$ as a function of magnetic field. (c) The temperature dependence of Hall mobility (blue circles). The black solid lines shows the power-law fitting results.

FIG. 3 (a) Temperature-dependent MR in the perpendicular magnetic field. (b) The $\Delta G$ curves at a series of temperatures (color circles). The HNL fitting lines are presented in the black solid lines. (c) The fitted values of $L_\varphi$ at the temperature regime 2 K − 50 K. The power-law fit shows $L_\varphi \propto T^{-3.3}$ and $\propto T^{-0.1}$ in the temperature regime 50 K - 25 K and 25 K - 2 K, respectively. The inset displays the dependence of $\alpha$ on the temperature.

FIG. 4 The curves of magnetoconductivity vs the perpendicular component of magnetic field. The inset shows the schematic of measurement where $\theta$ suggests the angle between the magnetic field and the current flow on the sample surface.

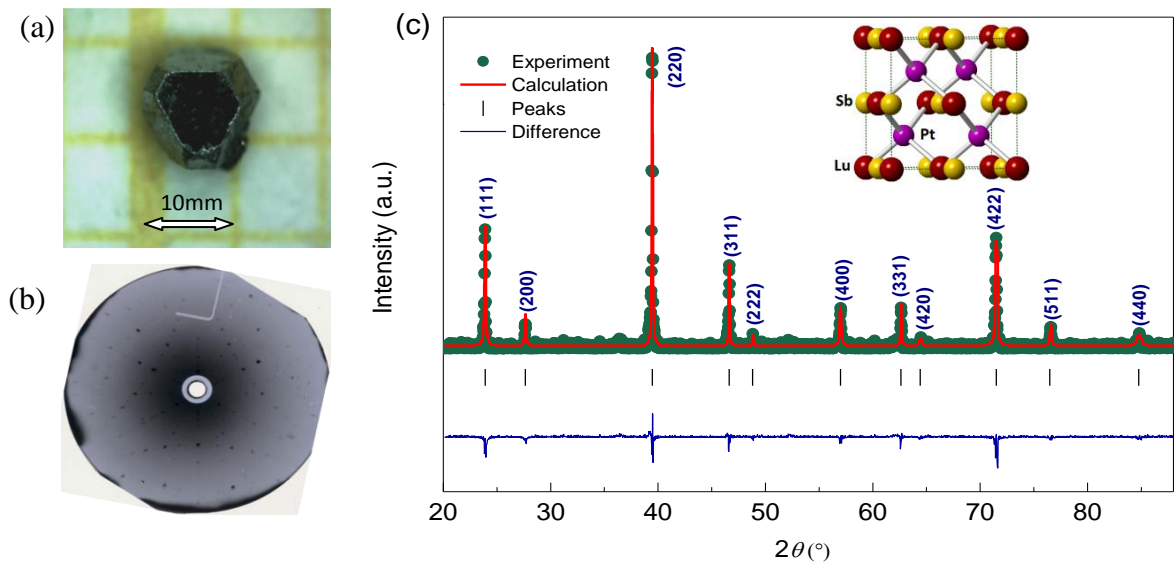

**FIG. 1 Hou et al.**

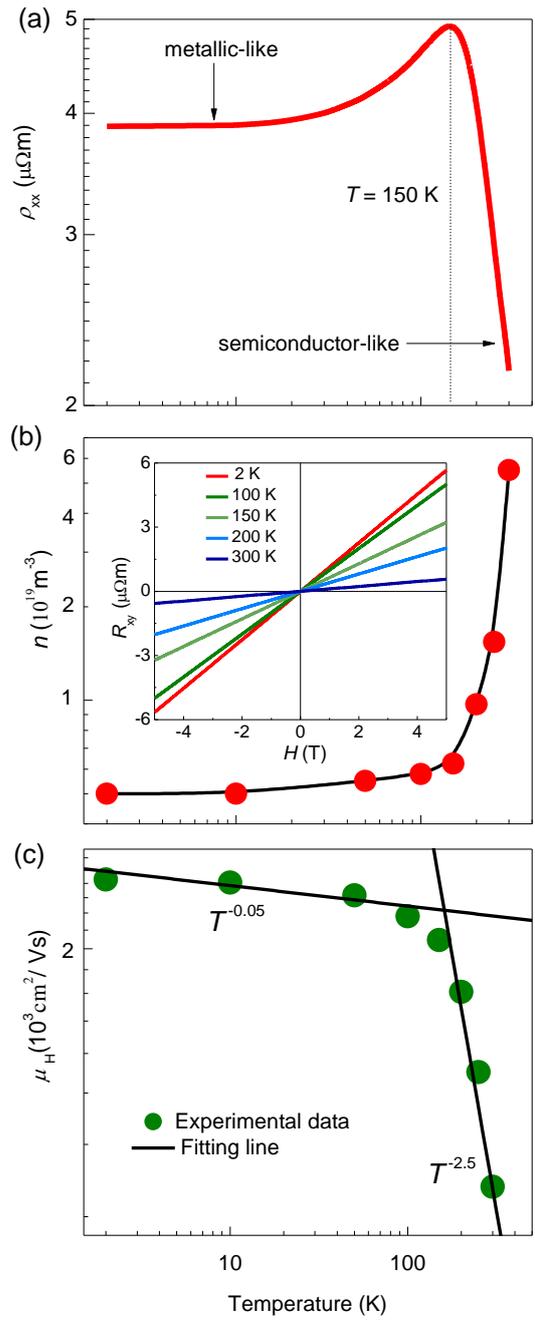

**FIG. 2 Hou et al.**

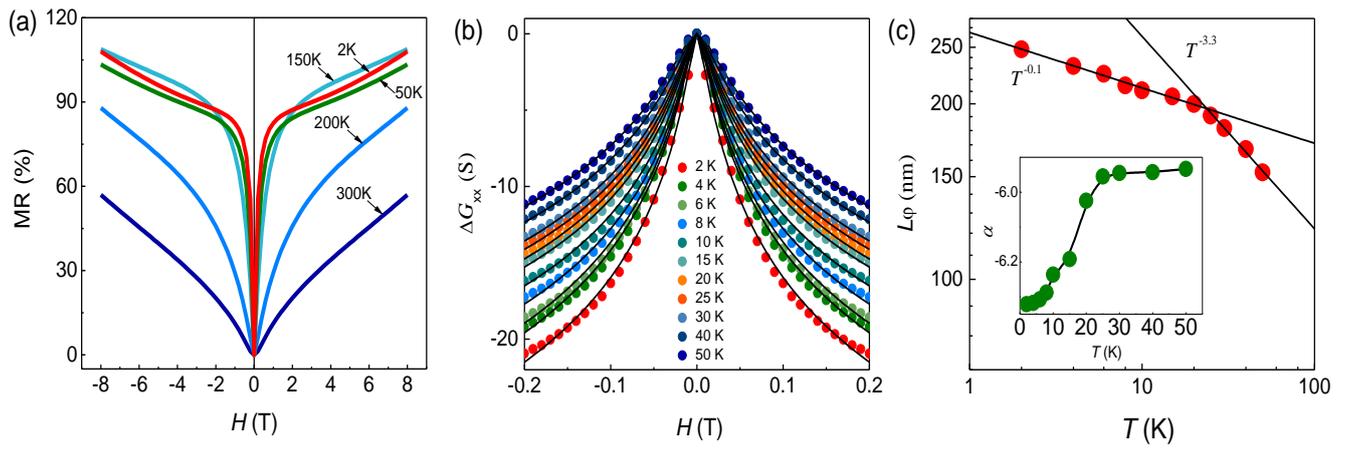

**FIG. 3 Hou et al.**

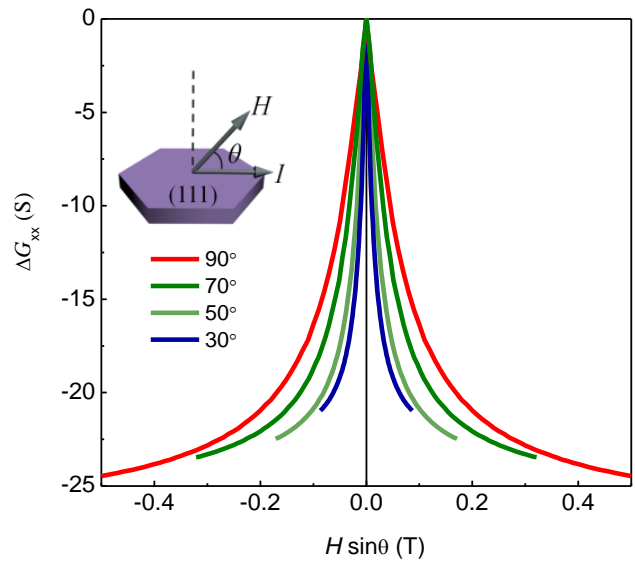

**FIG. 4** Hou et al.